\documentclass[12pt]{article}
\usepackage{epsfig}
\begin{document}
%
\begin{center}
{\Large {\bf Hybrid Baryons, a brief review}}
\footnote{Invited contribution to the COSY Workshop on Baryon Excitations,
J\"ulich, 2-3 May 2000.}
  \\[\baselineskip]
{\large T.Barnes}
  \\[\baselineskip]
{\large{\it Physics Division, Oak Ridge National Laboratory}}
  \\[\baselineskip]
{\large{\it Department of Physics and Astronomy, University of Tennessee}}
  \\[\baselineskip]
{\large{\it Institut f\"ur Kernphysik, Forschungszentrum J\"ulich}}
  \\[\baselineskip]
{\large{\it Institut f\"ur Theoretische Kernphysik, Universit\"at Bonn}}
  \\[2\baselineskip]
\end{center}
\vspace*{2.0cm}

\noindent
{\bf Abstract:}\\
{\small
This contribution is a brief review of the status of hybrid baryons, 
which are hypothetical baryons that incorporate 
a gluonic excitation. We first summarize the status of hybrid
{\it mesons}, since this closely related topic has 
seen considerable recent activity 
with the identification 
of two exotic candidates. Next we review theoretical expectations for the
masses and quantum numbers of hybrid baryons, which have come from studies of
the bag model, QCD sum rules and the flux tube model. 
Finally hybrid
baryon experiment is discussed, including suggestions for
experimenters at COSY.
}
\normalsize \\[\baselineskip]

\section{Hybrid Mesons}
Hybrids are hadrons in which the dominant component of the state 
consists of
quarks (antiquarks) and excited glue.
This 
deliberately 
vague definition is necessary at present because the subject has been studied
mainly through various models, and each assumes a particular description
of excited glue. Fortunately the models often reach similar 
conclusions regarding the quantum numbers and approximate masses of these
states, so from the experimental viewpoint there are relatively 
clear predictions regarding how one might find states with excited 
glue in the spectrum.

Hybrid mesons are usually modelled as $q\bar q$+excited glue, and this system 
can have all 
J$^{PC}$ 
quantum numbers. This implies that one can most usefully search for hybrids as
mesons with the so-called ``exotic quantum numbers"
J$^{PC} = 0^{--}, 0^{+-}, 1^{-+}, 2^{+-}, 3^{-+} \dots$ because these are
strictly forbidden to the nonrelativistic quark model's $q\bar q$ states.
Thus if one discovers a J$^{PC}$ exotic meson, it is certain that something 
other than $q\bar q$ has been found. Whether this is a hybrid or not may be a
more difficult question to answer, and depends on a comparison with theoretical
predictions for the masses, quantum numbers and decay couplings of hybrids.

Various models and methods have been used to predict the spectrum of hybrid
mesons, including the bag model, QCD sum rules, the flux tube model, 
and lattice gauge theory. All these approaches agree that there should be
a light exotic J$^{PC} = 1^{-+}$ hybrid, with a somewhat model dependent mass 
from {\it ca.}~1.5~GeV (bag model) to 1.9~GeV 
(flux tube model) to 2.0~GeV (LGT).
Only the flux tube model has been studied in detail for its decay mode 
predictions, and it leads to an {\it a priori} 
rather surprising expectation that
hybrid mesons decay preferentially to meson pairs with one internal 
orbital excitation, the so-called ``S+P'' modes, 
for example $\pi f_1$ and $\pi b_1$ 
rather than the more familiar ``S+S'' modes
such as $\pi\pi$, 
$\pi\eta$ and
$\pi\rho$. 

Experimentally we now have two candidate J$^{PC}$ exotic mesons, with
the same quantum numbers, a rather broad
I=1 $1^{-+}$ 
$\pi_1(1400)$ seen in $\eta\pi$ \cite{pi1_1400}, and a
somewhat narrower 
$\pi_1(1600)$ seen in $\rho\pi$, $\eta\pi$ and $b_1\pi$ \cite{pi1_1600}.
Although it is exciting to have {\it possible} confirmation of the existence
of exotic hybrid mesons, 
it is disturbing that the masses of these states are rather
lower than the {\it ca.}~2~GeV 
expected by LGT and the flux-tube model, and their
observation with large partial widths in the S+S modes $\pi\eta$ and
$\pi\rho$ disagrees with expectations of the flux-tube decay model.
Although there have been speculations that the $\pi_1(1400)$ 
in particular might be a nonresonant final-state effect, there have been
no model calculations that show this possibility is viable.
So, we are faced with a somewhat ambiguous situation, in which hybrid mesons
{\it may} have been discovered, but there are important disagreements with
theoretical expectations for masses and important decay modes.

\section{Theoretical Expectations for Hybrid Baryons}

\subsection{General expectations}

Augmenting the quarks $q$ and antiquarks $\bar q$ by gluons $g$ 
leads to additional 
states in the spectrum relative to the expectations of the
naive $q\bar q$ and $qqq$ quark model.
Physically allowed (color singlet) states in the baryon spectrum may be
constructed from $|qqqg\rangle$ ``hybrid'' basis states, in addition to the
familiar $|qqq\rangle$ quark model states:

\begin{displaymath}
|qqq\rangle\bigg|_{color}  = {\bf 1}\otimes 8 \otimes 8 \otimes 10 \ ,
\end{displaymath}
\begin{displaymath}
|qqqg\rangle\bigg|_{color} = 
\bigg(1\otimes 8 \otimes 8 \otimes 10\bigg)\otimes 8 
=  {\bf 1}^2 \otimes 8^5\otimes ... \ .
\end{displaymath}
The lowest hybrid baryon basis state is 
color octet and spatially symmetric
in the $qqq$ part of $|qqqg\rangle$,
making it a ${\bf 70}$ of SU(6). Since this
$qqq$ subsystem is combined with the angular momentum of the gluon,
we find the interesting prediction that 
{\it $|qqqg\rangle$ 
multiplets do not span the
same $|{\rm flavor},{\rm J}_{tot}\rangle$ states as an SU(6) $|qqq\rangle$ 
multiplet}. Thus we should find
evidence for ``incomplete'' or ``overcomplete'' SU(6) baryon multiplets
due to the presence of hybrids. 
More detailed predictions for the multiplet content typically
require the use of a specific model, although there has recently been
work on the derivation of 
general properties of hybrid baryon states and their decays in the
large quark mass and large-$N_c$ limit \cite{Cho99}.

\subsection{Bag model}

This model places relativistic quarks and gluons in a spherical cavity and
allows them to interact through QCD forces such as one gluon exchange (OGE),
the color Compton effect and so forth. Incorporation of these interactions
to leading nontrivial order, $O(\alpha_s)$, gives predictions for the 
spectrum and Hilbert space decomposition 
of light hybrid baryons in this 
model.
\vskip 0.5cm

\begin{figure}[ht]
\epsfig{file=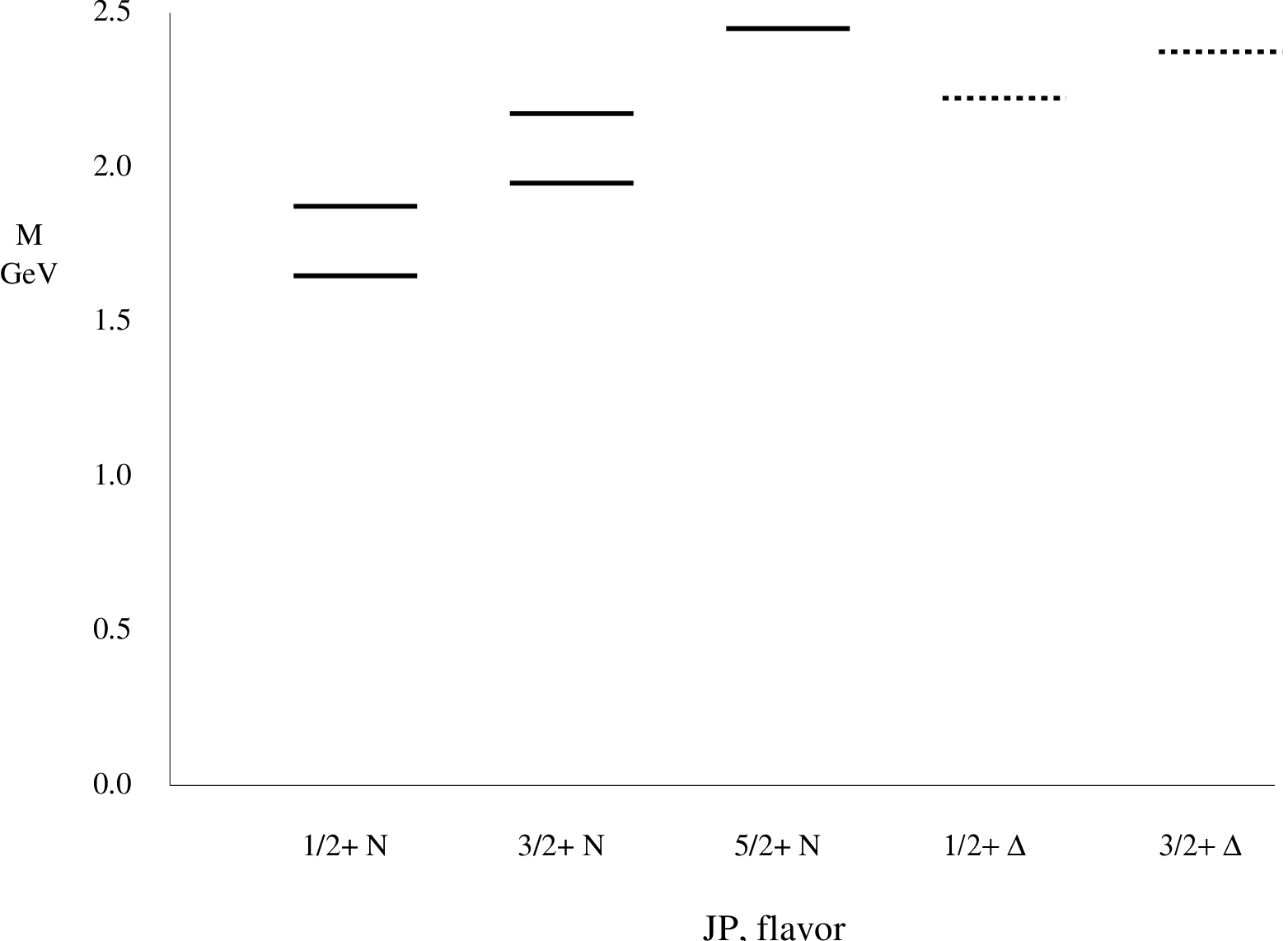,width=12.0cm}
\caption{\small{
The spectrum of light nonstrange hybrid baryons found by Barnes and Close
\cite{Bar83a} in the bag model.
}}
\end{figure}

The first published calculation of light bag model hybrids 
was due to
Barnes and Close \cite{Bar83a}, who derived the spectrum of
nonstrange $nnng$ states. 
This reference found the spectrum 
shown in Fig.1; in order of increasing mass the states are
\begin{displaymath}
\big(
1/2^+{\rm N} 
\big)^2 
\ ;
\big(
3/2^+{\rm N} 
\big)^2 
\ ;
\big(
1/2^+\Delta
\big) 
\ ;
\big(
3/2^+\Delta
\big) 
\ ;
\big(
5/2^+{\rm N} 
\big)
\ .
\end{displaymath}
Note that the lightest 
hybrid baryon is predicted to be an ``extra'' $1/2^+$
N$^*$ P$_{11}$ state 
(``extra'' meaning an overpopulation relative to the 
predictions of the $qqq$ quark model)
at about 1.6~GeV, which might possibly be identified with 
the Roper resonance. A subsequent calculation by Golowich, Haqq and Karl
\cite{Gol83} basically confirmed these results, but used a parameter 
set that gave a mass of about 1.5~GeV for this lightest hybrid, so 
identification with the Roper was given more support.
Carlson and Hansson \cite{Car83} extended these studies to strange hybrid
baryons, and found that the bag model predicted two relatively light 
$uds$-flavor hybrids, with the lightest expected at M$(1/2^+) = 1.63(4)$~GeV
with their parameters.

Decays are not usually considered in bag model calculations, which 
normally assume 
that the 
states are stable Hamiltonian eigenstates. A model of decays 
of bag model 
hybrid baryons was developed by Duck and Umland
\cite{Duc83}, who studied the 
$\Delta\pi$ and
N$\pi$ decay modes of the lightest N-flavor hybrid states. They concluded 
that the lightest hybrid has a much larger coupling to $\Delta\pi$ than N$\pi$.

\subsection{QCD sum rules (and LGT)}

This approach, which has been applied to hybrid baryons in several papers
of the previous decade, 
finds the masses and other parameters of the lowest-lying states 
in terms of numerically known VEVs, called ``condensates''. Since the sum rules
relate known VEVs to a sum of resonance contributions, there are systematic
uncertainties in separating the individual resonance and ``continuum'' 
parts. Identification of excited states such as hybrids is rather
difficult in this approach, since higher-mass contributions to 
the sum rules are suppressed exponentially. This exercise can be carried
out for hybrids, for example by calculating matrix elements of 
several operators and diagonalizing the result. In practice
the calculations also use $qqqg$ operators, which one would expect to
have larger hybrid couplings. To date only hybrids in the nucleon/Roper
sector $1/2^+$N have been studied using QCD sum rules. 

The first published hybrid baryon QCD sum rule calculation was by 
Martynenko \cite{Mar91}, who estimated the lightest $1/2^+$N hybrid baryon
mass to be near 2.1~GeV. A subsequent study by Kisslinger and Li
\cite{Kis95} reported algebra errors in the (very intricate)
matrix elements calculated in the Martynenko paper, and published 
a revised mass estimate of about 1.5~GeV, again very suggestive of the Roper.
A more recent review by Kisslinger \cite{Kis98} concluded that the Roper is
largely a hybrid (meaning dominantly $|qqqg\rangle$), the nucleon has little
evidence for a hybrid component, and also considers how one might 
calculate strong couplings. Some of this program of decay calculations
was carried out by Kisslinger and Li \cite{Kis99}, who conclude that the
lightest hybrid should have a rather small branching fraction 
ratio N$(\pi\pi)_S$ / N$\pi$, consistent with observation for the Roper.

We note in passing that lattice gauge theory uses a very similar
technique to QCD sum rules for extracting masses, and can also
be used to determine hybrid baryon masses, for example  
as nonleading contributions
to baryon operator correlation functions. Hybrids are not specifically
identified 
as such, but will appear in any determination of excited baryon masses
in LGT. In future, couplings to $qqq$ versus $qqqg$ operators may 
identify states with large excited gluonic components.  
Preliminary results for the first $1/2^+$ N$^*$ resonance in LGT
have been reported by Sasaki {\it et al.} \cite{Sas00}, and it appears
that this 
excited baryon may be identified through the use of a baryon
operator that has little overlap with the nucleon. This approach works well
for heavier quarks, but has not yet been carried out 
with high statistics for light quark masses.

\subsection{Flux tube model}

The flux tube model assumes that glue forms a dynamically excitable tube 
between quarks and antiquarks, and that the lightest 
hybrids are states in which this 
flux tube is spatially excited. The model is of special interest because 
of the reasonable agreement between its predictions for the mass of the
lightest J$^{PC}$-exotic meson and the presumably more reliable LGT
prediction ({\it ca.} 1.9~GeV in the flux tube model, 
versus {\it ca.} 2.0~GeV from LGT). The determination of excited states
in the baryon sector ($qqq$ + spatially excited flux tube) is a rather 
complicated problem which has only recently been treated. 
Flux tube model predictions for the lightest hybrid baryons were 
reported by Capstick and Page in 1999 \cite{Cap99}. They find that the
lightest hybrid baryons in the $nnn$ flavor sector are two each of
$1/2^+$N and $3/2^+$N, all at a mass of 1.87(10) GeV. Their results for
the N flavor hybrids are shown in Fig.2, together with current experimental
data.

\begin{figure}[ht]
\hskip 2.0cm
\epsfig{file=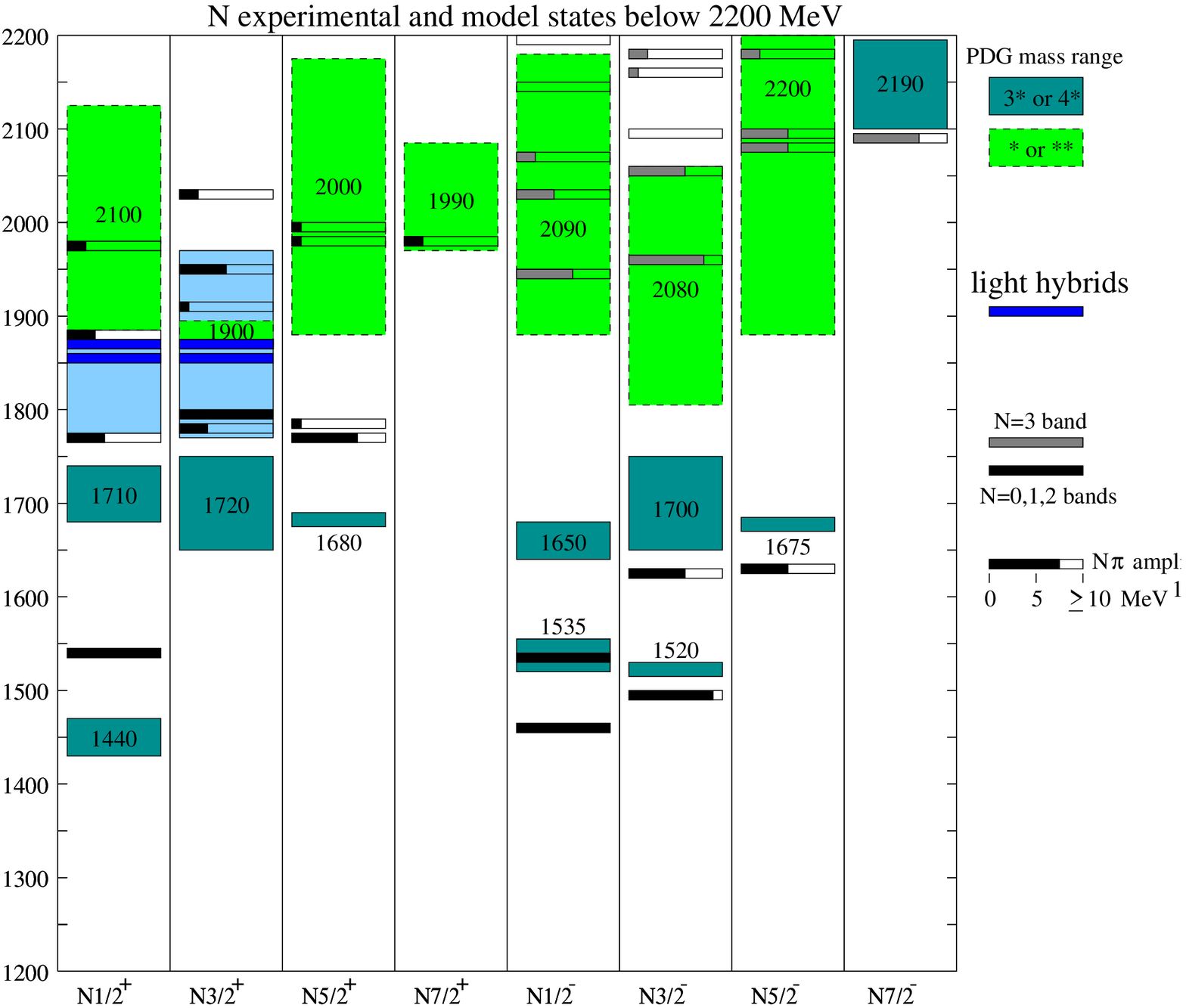,width=12.0cm}
\caption{\small{
The four light N flavor hybrid baryons found by Capstick and Page
\cite{Cap99} in the flux-tube model, compared to experiment. The
estimated error is $\pm 100$~MeV (light blue background).
}}
\end{figure}

Thus the lowest flux-tube hybrid baryon level is
predicted to include Roper quantum numbers, as was 
found in the bag model, {\it albeit} twofold degenerate and at a 
higher mass. In addition a 
degenerate $3/2^+$N pair is
expected. There are other differences in the multiplet content; 
the flux-tube hybrid baryon multiplet contains the
states
\begin{displaymath}
\big(
1/2^+{\rm N} 
\big)^2 
\ ;
\big(
3/2^+{\rm N} 
\big)^2 
\ ;
\big(
1/2^+\Delta
\big) 
\ ;
\big(
3/2^+\Delta
\big) 
\ ;
\big(
5/2^+\Delta
\big) 
\ ,
\end{displaymath}
so the flux tube model finds a
high-mass $5/2^+\Delta$; in the bag model the 
high-mass state was a $5/2^+$N. The flux-tube $\Delta$ hybrids are 
predicted to be degenerate, with a mass of 2.09(10)~GeV.

The implications of this work for searches for hybrid baryons, including 
various experimental search strategies such as overpopulation, strong decays,
EM couplings and production amplitudes, were recently reviewed by Page 
\cite{Pag99_00}. In particular Page suggests searches for hybrids 
in the final states 
N$\eta$,
N$\rho$ and 
N$\omega$.

Although there are no strong decay amplitude calculations reported for 
flux-tube hybrid baryons as yet, one can see that the qualitative 
arguments that are applicable to flux-tube hybrid mesons should apply here
as well. Thus one would expect the flux-tube decay model to predict 
that the largest couplings are to hadrons with internal orbital 
excitation. Phase space would clearly prefer the orbital excitation to be in 
the baryon, so hybrid baryons decays to $\pi S_{11}(1535)$ for example may be
favored. Since the 
$S_{11}(1535)$ 
has a large and characteristic 
N$\eta$
branch, study of the decay chains
\begin{equation}
hybrid \to \pi  S_{11}(1535) \ ; S_{11}(1535) \to {\rm N}\eta 
\end {equation}
and
\begin{equation}
hybrid \to \eta  S_{11}(1535) \ ; S_{11}(1535) \to {\rm N}\eta 
\end {equation}
may reveal hybrids in final states in which conventional baryons have somewhat
suppressed couplings. The chain
$hybrid \to \eta  S_{11}(1535) \to \eta\eta $N is especially attractive 
for detectors with good photon detection; such a study is planned at ELSA
using the Crystal Barrel detector \cite{Bac}.

\section{Identifying Hybrid Baryons}

\subsection{General Strategies}

Hybrid {\it mesons} 
can be studied most easily by searching for J$^{PC}$ exotics;
although we cannot be certain that a J$^{PC}$ exotic {\it is} a hybrid,
we can be certain that it is not a $q\bar q$ state. If 
such an exotic is found, one can then search for the exotic and non-exotic
partners that would confirm the presence of a hybrid multiplet.

In baryons there are unfortunately no J$^{P}$ exotics, so we must use other
properties of baryons to determine whether or not they are hybrids. In this
approach we classify the ``background'' of $qqq$ states, learn to describe
their couplings accurately, and then identify non-$qqq$
baryons through unusual couplings or as an overpopulation of states relative
to the $qqq$ quark model predictions.
One property of baryon resonances often cited as a hybrid discriminator
is their EM couplings, in other words their photoproduction and 
electroproduction amplitudes. Other possible ways to identify hybrids 
are through anomalous strong 
decay amplitudes and production systematics in novel channels such
as $J/\Psi$ hadronic decays. 

\subsection{Photocouplings}

Since photocouplings of baryon resonances are widely considered to be 
reasonably well predicted by the $qqq$ quark model, and new experimental 
facilities such as Jefferson Lab will provide detailed results on baryon
resonance photocouplings, the EM couplings predicted for hybrid baryons have
received considerable attention.

Photocouplings of bag model hybrid baryons were derived by Barnes and Close
\cite{Bar83b}; those of the lightest ``Roperlike'' hybrid were of special 
interest due to problems with predicting the EM couplings of the Roper in the
$qqq$ quark model. Barnes and Close found a generalization of the 
Moorhouse 
selection rule \cite{Mor66}, 
which in this case gave a vanishing photocoupling of the
I$_z=+1/2$ hybrid state from a proton,
\begin{equation}
\gamma {\rm P} \ \not  \!\!\; \to \ {\rm P}_g \ .
\end{equation}
Experimentally this photocoupling is {\it not} small, which apparently 
invalidates the identification of the Roper with this bag model hybrid.
Caution is appropriate here. First, the photocouplings are due to the basis
transitions
$\gamma |qqq\rangle  \to |qqq\rangle$ 
and
$\gamma |qqqg\rangle \to |qqqg\rangle$,
so $\gamma {\rm P} \to {\rm P}_g $  
actually tests the non-valence components
$|qqq\rangle$ in the hybrid and $|qqqg\rangle $ in the nucleon.
These nonleading amplitudes may be strongly model dependent!

Subsequently it was noted by 
Li \cite{Li91} that whereas Barnes and Close had 
assumed that the basis state 
$|^{2S_{qqq}+1}{\rm flavor}_{glue}\rangle = 
|^4{\rm N}_g\rangle$ 
was the zeroth-order ``Roper'' hybrid basis state, as suggested
by cavity QCD perturbation theory, one should actually use 
degenerate perturbation theory in this mixing problem because 
at zeroth order the 
basis states
$|^2{\rm N}_g\rangle$ and $|^4{\rm N}_g\rangle$ are degenerate.
The Moorhouse selection rule does not apply to this 
$|^2{\rm N}_g\rangle$ component,
so one might still identify the Roper with the bag model hybrid if this is 
indeed an
important configuration.

Carlson and Mukhopadhyay \cite{Car91} noted that electroproduction amplitudes
of baryon resonances with dominant hybrid
components
might be very characteristic. 
They concluded that the transverse-photon 
$qqqg$ 
electroproduction form factor
should fall faster than 
the 
corresponding 
$qqq$ form
by an additional factor of $1/Q^2$. 
Thus a rapid fall of electroproduction 
amplitudes with increasing $Q^2$ is a possible hybrid signature.

This suggestion was considered by Li, Burkert and Li \cite{Li92}, who
compared theoretical models for the $Q^2$ dependence of
electroproduction
amplitudes of
radial-$qqq$ and hybrid states
with experiment. 
In their study the 
theoretical radial-$qqq$ electroproduction amplitude was quite large,
and unlike the hybrid electroproduction amplitude 
did not fall rapidly with $Q^2$.
They
concluded that the rapid fall of the experimental Roper electroproduction
amplitude with increasing $Q^2$ favored the identification of the Roper with a
hybrid 
(see Fig.3).

\begin{figure}[ht]
\hskip 1.5cm
\epsfig{file=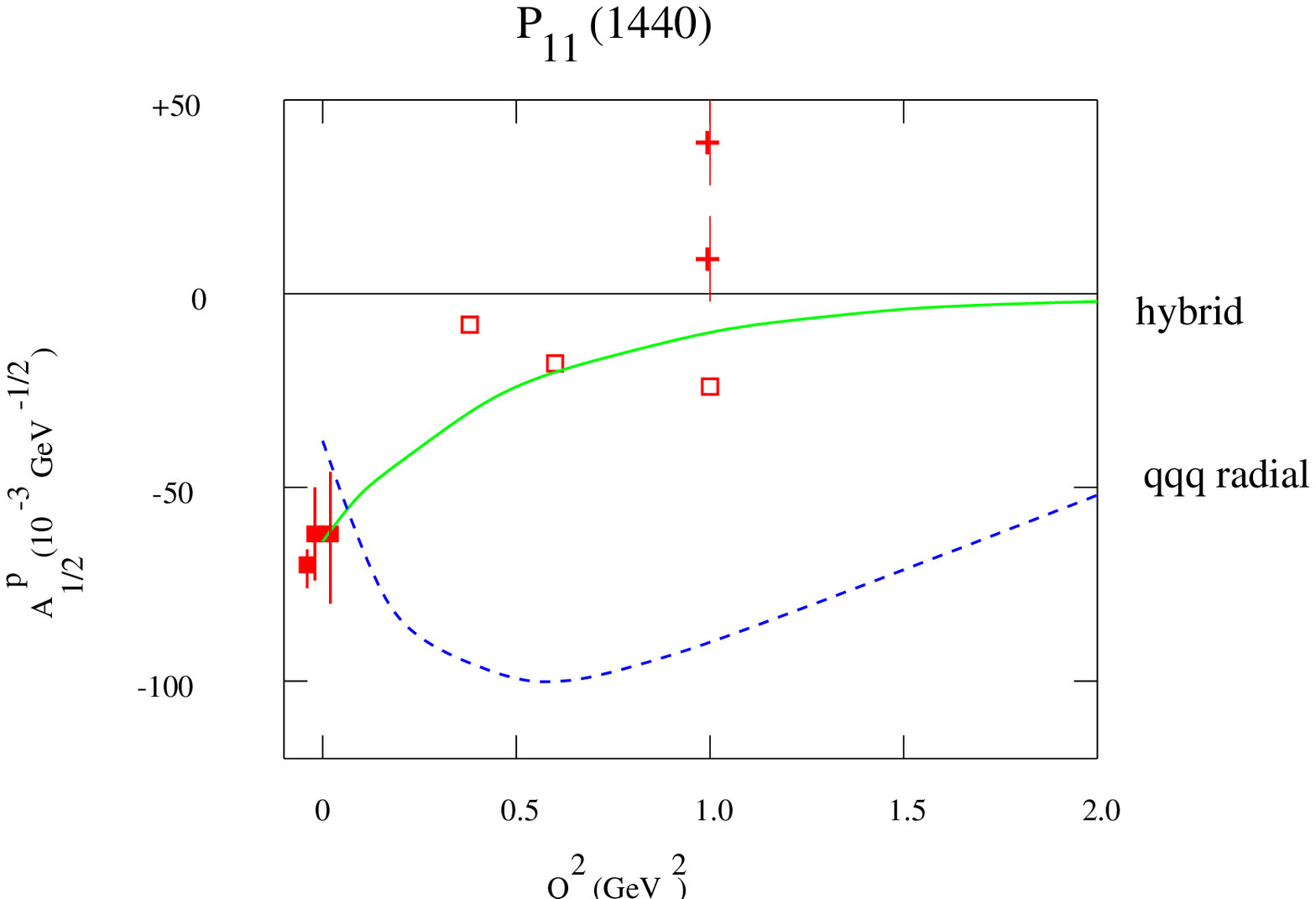,width=11.0cm}
\caption{\small{
An early comparison of hybrid and radial-$qqq$ electroproduction
amplitudes with experimental Roper results (from \cite{Li92}).
}}
\end{figure}

Unfortunately 
for this simple picture, subsequent calculations have shown that the amplitude
$\gamma (qqq)_{1S} \to (qqq)_{2S} $ is sensitive to the details of the
calculation, and small radial-$qqq$ electroproduction amplitudes 
can also be accommodated.
Although electroproduction shows great promise as a way to identify anomalous 
baryon resonances, until such time as
electroproduction calculations 
are shown to be reliable for a wide range of $qqq$ states, including radial
excitations, 
the classification of resonances through their EM couplings will be 
problematic.

\subsection{$J/\Psi$ hadronic decays.}

Rather surprisingly, BES at BEPC is being used to 
study N$^*$ spectroscopy using
$J/\Psi$ hadronic decays. Zou {\it et al.}  \cite{Zou99_00} note that
one might expect hybrid baryons to have larger production amplitudes from
$J/\Psi$ than conventional $qqq$ baryons, because a
$ggg$ state produced in $J/\Psi$
annihilation should have a larger overlap with a final hybrid baryon
(see Fig.4). It is certainly 
interesting to establish which baryons are produced with large amplitudes
in $J/\Psi$ annihilation, as any unusual states thus produced are possible
hybrid baryon candidates.

\begin{figure}[ht]
\begin{center}
\epsfig{file=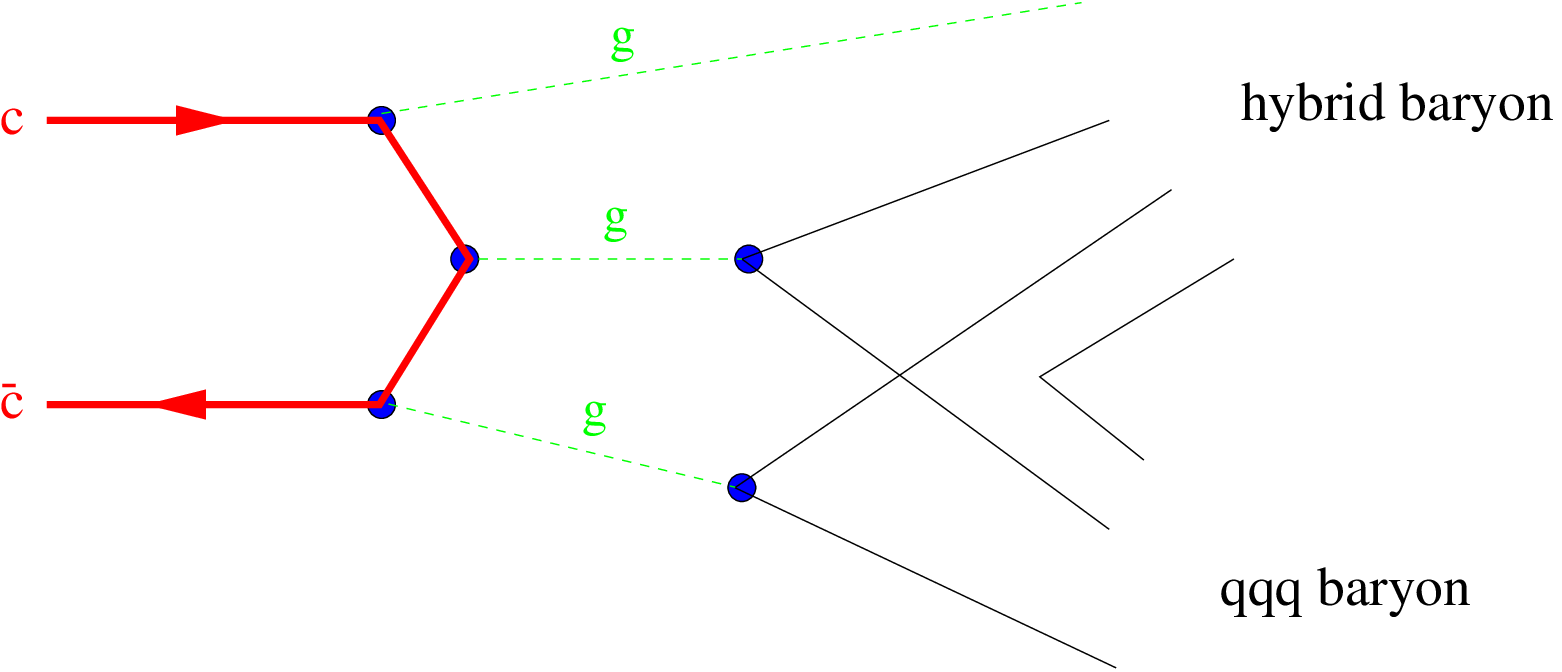,width=10.0cm}
\caption{\small{
Production of $qqqg$ states from 
$J/\Psi$ 
radiative decays occurs
at $O(\alpha_s^5)$ (followed by nonperturbative pair production), 
which leads
$J/\Psi \to ggg \to (qqq)+(\bar q \bar q \bar q )$
by
one power
of $\alpha_s$.
}}
\end{center}
\end{figure}

\begin{figure}[ht]
\hskip 1.5cm
\epsfig{file=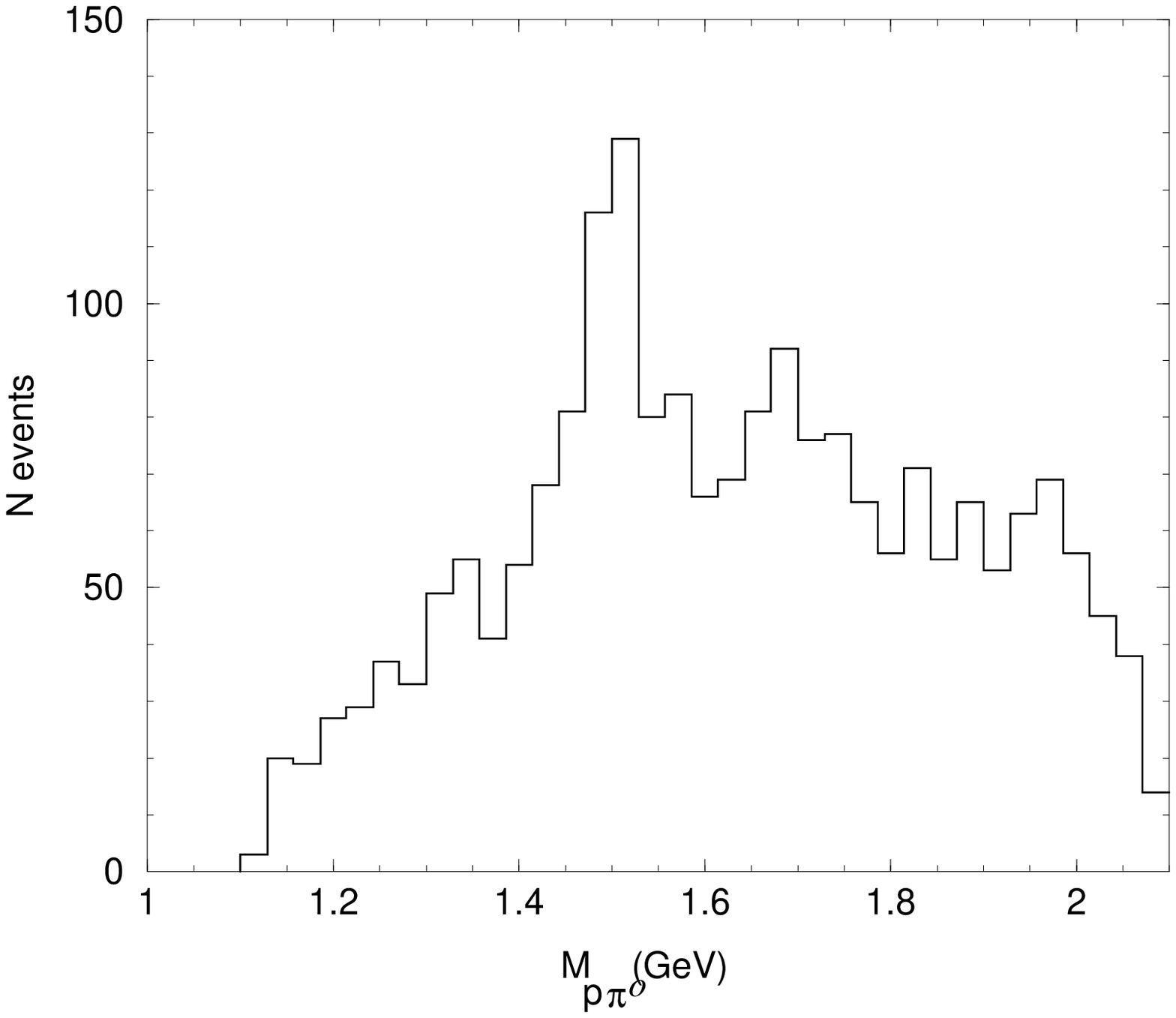,width=11.0cm}
\caption{\small{
BES data for the $p\pi^o$ invariant mass distribution in
$J/\Psi \to p p \pi^o$ \cite{Zou99_00}, showing evidence for the
$1/2^-$ N$_{11}(1535)$.
}}
\end{figure}

To date BES has 7.8M $J/\Psi$ events, from which they select
$p\bar p\pi^o$ 
and
$p\bar p\eta$. 
This approach has the additional advantage that it is an I$=1/2$ filter,
so the many $\Delta$ (and hybrid $\Delta$) states will not be present
to complicate the analysis. 
The only clear peak in the present data is the $S_{11}(1535)$, 
in the $p\bar p\pi^o$ channel (see Fig.5).  Since {\it ca.}~50M 
$J/\Psi$ 
events are expected in the near future, hybrid baryon candidates may yet be
identified in this process.

Another interesting possibility, which may be undertaken at COSY,
is the selection of isospin states through the choice of unusual beams.
This might be possible for example 
using an $\alpha$ beam, as discussed at this 
meeting by Morsch and collaborators \cite{Morsch}. This technique has
previously been shown to be effective in enhancing the Roper signal, and
could be useful in a search for the light 
N-flavor hybrids because the 
production of $\Delta$ states may be much weaker 
than in the more familiar
$\pi$N and $\gamma$N reactions.

\subsection{Overpopulation of the spectrum; strong decays}

Since there are no J$^P$ exotic 
baryons, searches for baryon hybrids are in effect
seaches for evidence of {\it overpopulation}, in which one attempts to 
establish whether there is clear experimental evidence for more states in
the spectrum than predicted by the $qqq$ quark model alone. 

This suggests that establishing the conventional $qqq$ baryon spectrum
and studying the properties of these ``ordinary hadrons'' is a very important
part of the search for exotica; what is unusual may only be evident once the 
background of conventional states is very well understood. This includes
not only the quantum numbers, masses and widths of the baryons, but also their
EM couplings and strong decay amplitudes, since these may be useful as 
probes of the internal structure of unusual resonances. This program of 
establishing {\it all} nonstrange 
baryon resonances should be pursued at least to 
{\it ca.}~2.2~GeV, since this is somewhat above the highest mass estimate 
for the lightest nonstrange hybrid baryon. This conservative strategy of
identifying all states is especially appropriate because we have little 
evidence regarding which model of hybrids is most accurate, and this must be 
decided by comparing with a reasonably complete experimental spectrum. 
In the worst case there could be large mixing angles between 
conventional and 
hybrid baryons, so that the distinction between these states 
is rather artificial; simple state counting is then the most direct
approach for establishing the presence of additional degrees of freedom.
Another complication is that
there are probably additional types of baryons, such as $(qqq)(q\bar q)$ 
``molecular'' states, which would presumably also be found in a detailed 
experimental study of the baryon spectrum.

Strong decay amplitudes may also prove useful in identifying
hybrid baryons, so the careful determination of 
strong branching 
fractions and decay amplitudes of experimental baryon 
resonances is especially 
important. In the meson case there are striking predictions that hybrids should
decay preferentially 
to states with internal orbital excitation, and if this rule is confirmed
we have a very useful signature that may also be applicable to
hybrid baryons. 
(This is being investigated by Black and Page \cite{Bla00}.)
One should note however that the strong decays of conventional $qqq$ baryons
may not be well understood; the usual models simply assume $q\bar q$ pair
production with $0^{++}$ quantum numbers, which appears to work well in 
certain test cases but has not been compared to high-statistics 
experimental resonance data for a wide range of 
states, simply because such data has
not been available. Since very detailed predictions for baryon
resonance decay amplitudes have now been published by Capstick and Roberts 
in this model \cite{Cap00} 
and data from CEBAF is becoming available, we should soon be able to
establish whether 
we can
predict conventional $qqq$ baryon strong decays accurately. If so, 
hybrid baryons and other exotica might be identifiable through their
anomalous strong decays.

\section{Summary and Conclusions}

An {\it overpopulation} of baryon resonances 
is expected relative to the predicitons of the $qqq$ quark model,
due to excitation of the glue degree of freedom. Excited glue will 
lead to novel baryon resonances, which are known as ``hybrid baryons''.
Analogous J$^{PC}$ exotic meson hybrids 
may already have been identified
(the $\pi_1(1400)$ and $\pi_1(1600)$).
Theorists have derived the spectrum of light hybrid baryons and some
of their properties using various models, and expect that the lightest
hybrid will have $1/2^+$N ``Roper'' quantum numbers and a (rather model
dependent) mass of {\it ca.}~1.5-1.9~GeV. (It is amusing that this
range of masses is also predicted 
for the lightest exotic hybrid
meson.)
Strange baryon hybrids are also anticipated, and a relatively light
$1/2^+\Lambda$ hybrid is predicted.
Hybrids may be identifiable through distinctive decay modes
such as $\pi +$ P-wave baryon (similar to the S+P modes of meson hybrids)
and unusual photo- and electroproduction amplitudes.

A systematic search for these light hybrids
can be carried out by completing the study of N$^*$ spectroscopy
to {\it ca.}~2.2~GeV, and by accurately determining the decay amplitudes
and flavor partners of all observed states.  
COSY 
can contribute to the search for hybrids by helping to establish 
the baryon
resonance spectrum and strong decay amplitudes 
up to this mass scale, ideally in both nonstrange and strange 
sectors.

\section{Acknowledgements}

It was a great pleasure to contribute to the COSY Workshop
on Baryon Spectroscopy by presenting this material on hybrid baryons.
I am indebted to N.Black, S.Capstick and P.R.Page for several discussions
of their recent work on hybrid baryons, and to S.Capstick for providing Fig.2. 
This research was supported in part by the
DOE Division of Nuclear Physics,
at ORNL,
managed by UT-Battelle, LLC, for the US Department of Energy
under Contract No. DE-AC05-00OR22725, and by
the Deutsche Forschungsgemeinschaft DFG
at the University of Bonn and the Forschungszentrum J\"ulich
under contract Bo
56/153-1.

\end{document}